\documentclass[letterpaper]{IEEEtran} 

\usepackage{amsmath,amsopn,amssymb,stackrel}
\usepackage{graphicx,xspace,color,soul}
\usepackage{epsfig,subfigure}
\usepackage{longtable,multirow}
\usepackage{array,float}
\usepackage{cite,citesort}
\usepackage{url}
\usepackage{bm}

\usepackage{algorithm,algorithmicx,algpseudocode}

\usepackage{epstopdf}
\usepackage{stfloats}

\def\h{{\mathbf h}}

\def\w{{\mathbf w}}

\def\I{\mathbf{I}}

\def\bC{\mathbb{C}}

\def\cC{{\mathcal C}}

\def\cK{{\mathcal K}}

\def\cN{{\mathcal N}}
\def\cR{{\mathcal R}}

\begin{document}
\title{User Pre-Scheduling and Beamforming with Imperfect CSI 
in 5G Fog Radio Access Networks}

\author{
Nicolas Pontois,
Megumi Kaneko~\IEEEmembership{Senior Member,~IEEE},\\
Thi H\`{a} Ly Dinh, 
Lila Boukhatem~\IEEEmembership{Member,~IEEE}
\begin{small}
        \thanks{M. Kaneko and T.H.L. Dinh are with the
        National Institute of Informatics, 2-1-2, Hitotsubashi, Chiyoda-ku,
        Tokyo, Japan 101--8430 
        (e-mail: \{megkaneko,halydinh\}@nii.ac.jp).} 
        \thanks{N. Pontois and L. Boukhatem are with  
        LRI, Paris-Sud University,
        Orsay, France
        (e-mail: nicolas.pontois@telecom-paristech.fr, lila.boukhatem@lri.fr).}
\end{small}
}%
\maketitle
\vspace{0.1in}

\begin{abstract}
We investigate the user-to-cell association (or user-clustering) and beamforming design for Cloud Radio Access Networks (CRANs) and Fog Radio Access Networks (FogRANs) for 5G. CRAN enables cloud centralized resource and power allocation optimization over all the small cells served by multiple Access Points (APs). However, the fronthaul links connecting each AP to the cloud introduce delays and cause outdated Channel State Information (CSI). By contrast, FogRAN enables lower latencies and better CSI qualities, at the cost of local optimization.
To alleviate these issues, we propose a hybrid algorithm exploiting both the centralized feature of the cloud for globally-optimized pre-scheduling using outdated CSIs and the distributed nature of FogRAN for accurate beamforming with high quality CSIs. 
The centralized phase enables to consider the interference patterns over the global network, while the distributed phase allows for latency reduction.
Simulation results show that our hybrid algorithm for FogRAN outperforms the centralized algorithm under imperfect CSI, both in terms of throughput and delays\footnote{This work is  supported by the Grants-in-Aid for Scientific Research (Kakenhi) no. 17K06453 from the Ministry of Education, Science, Sports, and Culture of Japan, and by the CNRS-PICS project between LRI and NII.}. 
\end{abstract}

\begin{IEEEkeywords}
5G, CRAN, FogRAN, user clustering, beamforming, radio resource allocation
\end{IEEEkeywords}

\IEEEpeerreviewmaketitle

\section{Introduction}
\label{sec:intro}

The fifth generation (5G) communication system is expected to support the ever increasing demands for mobile data traffic under severe spectrum deficiencies, while satisfying more stringent user Quality of Service (QoS) levels. To achieve this, Cloud Radio Access Networks (CRANs) are considered as a key enabling technology, by incorporating cloud computing capabilities at the service of radio access~\cite{Che17sep}. 
A CRAN covers a large communication area divided into dense small cells served by Remote Radio Heads (RRHs), i.e., simple Access Points (APs) with only basic functionalities such as Radio Frequency (RF) and A/D conversion. In CRAN, signal processing and radio access tasks are performed in a centralized manner by the cloud Baseband Units (BBUs) forming a powerful server referred as a BBU pool. Signals of the mobile users in each small cell are transmitted between each AP and the cloud via the fronthaul links. 
Although this fully centralized architecture enables optimal joint baseband signal processing and radio resource allocation/interference management, it imposes heavy burden on the capacity/delay-limited fronthaul links. To cope with these issues, there has been tremendous interest for optimizing user clustering and beamforming under fronthaul constraints~\cite{Dai14oct}\cite{Dai15sep}.
Another major drawback of this centralized architecture is the additional network latency introduced by fronthaul links, making it unsuited for the highly delay-sensitive applications envisioned in 5G. Such delay also entails outdated Channel State Information (CSI) knowledge at the cloud side of the links between all APs and users, causing important performance degradation of resource allocation and beamforming schemes in CRAN~\cite{Wan17mar}.

Thus, recently there has been the advent of “moving the intelligence towards the edge", giving rise to Mobile Edge Computing (MEC) systems also known as Cloudlets or Fog Radio Access Networks (FogRAN)~\cite{Pen16jul}. Toward this end, FogAPs are now equipped with more functionalities compared to RRHs, e.g., computing and caching capabilities. This structure is expected to drastically alleviate the burden on fronthaul links and to meet the stringent delay requirements of edge users~\cite{Shi17jan}, but at the cost of lower network-wide optimality. Many works have exploited this edge processing to enhance the performance of various applications or analyzing the joint optimization of cloud/edge processing~\cite{Par16nov}.  
However, there have been few works  on the design of optimized physical/MAC layers under this novel FogRAN architecture, in particular regarding user clustering and beamforming issues. This is a crucial problem since optimized lower layers will have a huge impact on the actual performance of FogRANs at the application level.  

Therefore, in this work we investigate the joint user clustering and beamforming problem in FogRANs. We propose a resource allocation scheme that enables to exploit both the centralized processing capabilities of the cloud and the distributed computing features of FogAPs. It first carries out a centralized user pre-scheduling that provides the optimal user clustering to each FogAP, taking into account all interferences based on global but outdated CSI knowledge, similarly to the CRAN case. Then, the actual beamforming is computed at each FogAP for its own allocated users by pre-scheduling, using accurate CSI knowledge since the delay of CSI feedback is negligible compared to the transport delays due to fronthaul links. Our proposed scheme provides an optimized trade-off between centralized cloud processing for large-scale user clustering and distributed local beamforming, given heterogeneous CSI qualities.  
This is because user clustering is not that sensitive to CSI accuracy, unlike beamforming whose performance crucially depends on it. 
The numerical evaluations show that, compared to the reference centralized CRAN optimization, our proposed method provides similar sum-rate performance for much reduced latencies, in the presence of outdated CSIs due to fronthaul delays. 

\section{System Model}
\label{sec:system}
\subsection{CRAN/FogRAN architectures for core/edge intelligence}
\label{subsec:CRAN}

\begin{figure}[t]
\centering
\includegraphics[scale=0.32]{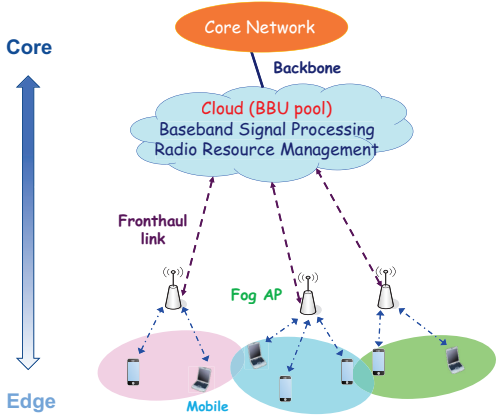}
\vspace*{-0.5cm}
\caption{CRAN/FogRAN architectures}
\label{fig:FogRAN}
\end{figure}

We consider two types of architectures referred as CRAN and FogRAN depending on the intelligence location either towards the core or edge, as depicted in Fig. \ref{fig:FogRAN}.
In the CRAN case, we assume a centralized system where all the signal processing and resource management tasks are performed at the cloud Baseband Unit (BBU) pool. $R$ macro or pico RRHs (APs) in set $\mathcal{R}$ are connected to the cloud through fronthaul links of respective capacities $C_r$. 
Each AP $r$ is equipped with $M_r$ transmit antennas. The set of all mobile users is denoted by $\mathcal{K}$ with cardinality $K$. Each user terminal is equipped with one antenna. 
We denote by $\w_{rk} \in \bC^{M_r \times 1}$ the beamforming vector of AP $r$ to user $k$. The concatenated beamforming vector of all AP antennas is defined as $\w_k = [\w_{1k}^H, \w_{2k}^H, \cdots, \w_{Rk}^H]^H \in \bC^{M \times 1}$ for user $k$, where $M = \sum_{r\in \cR}M_r$ is the total number of transmit antennas and $(.)^{H}$ denotes Hermitian transpose. Similarly, $\h_{rk} \in \bC^{M_r \times 1}$ is the channel vector between AP $r$ and user $k$ and $\h_k = [\h_{1k}^H,\h_{2k}^H, \cdots, \h_{Rk}^H]^H \in \bC^{M \times 1}$ the channel vector from all APs to user $k$. 
Then, the received signal $y_k$ by user $k$ is given by
\begin{equation}
	y_k = \h_k^H\w_{k}s_{k} + \h_k^H\sum_{{\substack{k' \in \cK\\ k' \neq k}}}\w_{k'}s_{k'} + n_k,
\label{eq:y}
\end{equation}
where $s_k$ is the transmit message for user $k$ drawn independently from the signal constellation with zero mean and unit variance, and $n_k \sim \cC\cN (0,\sigma_n^2)$ denotes the AWGN noise. The first term in (\ref{eq:y}) is the desired signal, and the second is the interference resulting from the other users' signals.
The beamforming vectors $\w_{k}$ will be optimized at the cloud BBUs for all users, and any user may be served by any of the $R$ APs.

%
%



In the FogRAN architecture, the intelligence is pushed towards the edge by enhancing traditional RRHs with higher processing capabilities, allowing basic signal processing tasks. Therefore, for differentiation these RRHs will be referred as FogAPs as in Fig. \ref{fig:FogRAN}.
In our proposed scheme, the beamforming vectors will be optimized locally at each FogAP $r$. The received signal of user $k$ served by FogAP $r$ is also given by (\ref{eq:y}), but where in $\w_k = [\w_{1k}^H, \w_{2k}^H, \cdots, \w_{Rk}^H]^H$, only the beamforming vector $\w_{rk}$ corresponding to the Fog AP $r$ associated to user $k$ is non-zero\footnote{The proposed scheme also works if each user is associated to more than one FogAP. Such intermediate solutions will be further explored.}. The set of users associated to FogAP $r$ is denoted $\mathcal{K}_r$ with cardinality $K_r$.

From (\ref{eq:y}), the Signal to Interference-plus-Noise Ratio (SINR) of user $k$ is written as
\begin{equation}
	\gamma_k = \frac{\left|\h_k^H\w_k \right|^2}{|\sum\limits_{\substack{k' \in \cK \\ k' \neq k}}\h_k^H\w_{k'}|^2 + \sigma_n^2}.
\label{eq:gammaG}
\end{equation}
The achievable rate for user $k$ is thus $\log(1+\gamma_k)$.
The Signal to Leakage-plus-Noise Ratio (SLNR) of user $k$ is defined as 
\begin{equation}
	\zeta_{k}= \frac{|\h_{k}^H\w_{k}|^2}{|\sum\limits_{\substack{k' \in \cK \\ k' \neq k}}\h_{k'}^H\w_{k}|^2 + \sigma_n^2},
\label{eq:zetaG}
\end{equation}
where in the denominator, we have the total power leakage towards all other users $k'$ from FogAP $r$ due to its signal transmitted to user $k$ with beamforming vector $\w_{k}$.

\subsection{Imperfect Channel State Information}
\label{subsec:imperfectCSI}

In centralized CRAN, optimal resource allocation can be performed using global CSI knowledge, i.e., all channel vectors $\h_{rk}$ for all APs $r$ and all users $k$. However, the fronthaul links will introduce non-negligible delays as pointed out in ~\cite{Pen16jul} causing outdated CSI.
The stochastic error model will be assumed as in \cite{Wan17mar}\cite{Du12mar}, where the imperfect channel vector is given by
 \begin{equation}
	\widetilde{\h}_{rk} = \h_{rk} + \mathbf{e}_{rk},
\label{eq:h}
\end{equation}
where $ \mathbf{e}_{rk} \sim \cC\cN(\mathbf{0},\sigma_e^2\I_{M_r})$. 
Then, the concatenated imperfect CSI is defined as $\widetilde{\h}_{k}=[\widetilde{\h}_{1k}^H, \widetilde{\h}_{2k}^H, \cdots, \widetilde{\h}_{Rk}^H ]^H \in\bC^{M \times 1}$.
Thus, only these outdated channels $\widetilde{\h}_{k}$ for all users $k$ will be available at the BBUs, i.e., \textit{global but imperfect CSI}. By contrast, in the FogRAN case, perfect knowledge of CSI $\h_{rk}$ will be assumed at each FogAP $r$, but only for its associated users and without any knowledge of interference channels, i.e., \textit{perfect but local CSI}.

\section{Reference Centralized algorithm for CRAN}
\label{sec:reference}



We focus on the weighted sum-rate maximization problem subject to fronthaul constraints as in \cite{Dai14oct}.
Optimal user clustering and beamforming vectors are determined at the BBU pool using global CSI. 
The optimization problem is formulated as 
\begin{equation}
	\max_{{\w}_{rk}} \sum_{k\in \cK}\alpha_kR_k
\label{eq:objCRAN}
\end{equation}
\begin{equation}
s.t. \sum_{k\in \cK_r}||{\w_{rk}}||_2^2 \le P_r, \quad \forall r \in \mathcal{R}
\label{eq:cons1}
\end{equation}
\begin{equation}
\sum_{k \in \cK_r}R_k \le C_r, \quad \forall r \in \mathcal{R}
\label{eq:cons2}
\end{equation}
\begin{equation}
R_k \le \log(1 + \gamma_k), \quad  \forall k \in \mathcal{K}
\label{eq:cons3}
\end{equation}
where $\alpha_k$ are weight parameters to achieve different fairness levels among users. 
The first constraint is given by the maximum power for each AP $r$, the second one is the per-AP fronthaul rate constraint, and the third one expresses the achievable rate for each user $k$.

This is a non-convex optimization problem for which a weighted MMSE-based algorithm was proposed \cite{Dai14oct}\cite{Dai15sep}. The case with perfect CSI represents the ideal scenario in terms of system performance, but requires full CSI feedback for all users from each AP, resulting into a significant burden over bandwidth-limited fronthaul links.
In reality, the CSI used for this optimization will be necessarily outdated due to the delays introduced by fronthaul links. Therefore, this reference algorithm for CRAN will be evaluated under different levels of CSI imperfectness.

\section{Proposed Hybrid Algorithm for FogRAN}
\label{sec:proposed}

In the proposed scheme, we split the joint resource allocation tasks: the user per-scheduling carried out centrally at the cloud BBUs, and the beamforming optimization carried out locally at each FogAP. The pre-scheduling consists in a user clustering, where the BBU pool decides to which FogAP each user should be assigned for given time frames. This pre-scheduling is performed periodically, every $T$ frames, based on outdated CSI due to fronthaul delays. Given the resulting user clustering, each FogAP performs beamforming in each frame, using perfect CSI.
Since FogAPs are uncoordinated during this beamforming phase, the pre-scheduling needs to determine optimal user clusterings forming a partition  $(\cK_r)_{r\in \cR}$ of the set of all users. This is in contrast with the CRAN user clustering in Section \ref{sec:reference}, where each user may be served by any AP. Note that some subsets $\cK_r$ may be empty, i.e., some FogAPs may not have any scheduled user for given frames.
 The details of each phase are given below.

\subsection{Pre-Scheduling}

For pre-scheduling, we solve a modified version of the weighted sum-rate optimization in CRAN (\ref{eq:objCRAN}), formulated as
\begin{equation}
	\max_{{\w}_{rk}} \sum_{k\in \cK }\alpha_k R_k
\label{eq:objFRAN}
\end{equation}
\begin{equation}
	s.t. \sum_{k \in \cK_r} ||\w_{rk}||_2^2 \le P_r, \quad \forall r \in \mathcal{R}
\label{eq:consFRAN1}
\end{equation}
\begin{equation}
	\sum_{k \in \cK_r}R_k \le C_r, \quad  \forall r \in \mathcal{R}
\label{eq:consFRAN2}
\end{equation}
\begin{equation}
	R_k \le \log(1 + \gamma_k), \quad  \forall k \in \mathcal{K}
\label{eq:consFRAN3}
\end{equation}
\begin{equation}
	\sum_{r\in \cR}||{\w}_{rk}||_0 = 1, \quad  \forall k \in \mathcal{K}
\label{eq:consFRAN4}
\end{equation}
where the last additional constraint enforces that each user is associated to only one FogAP, i.e., it ensures the partitioning mentioned above. This zero-norm constraint makes the optimization problem difficult by its discrete nature. To solve this problem, we use our solution in~\cite{Kat16aug} which is based on a relaxation technique similar to that in \cite{Dai15sep}. 
The obtained solutions give an implicit scheduling, so we can retrieve the user clustering as follows: $k \in \cK_r$ if ${\w}_{rk} \neq \mathbf{0}$.


\subsection{Local Beamforming}

In order to efficiently optimize the local beamforming, we propose to maximize the SLNR of each user at each FogAP. SLNR optimization is especially suited in this case since FogAPs are unable to coordinate among themselves and do not have access to the global SINR levels experienced by their associated users. In addition, this optimization requires very low complexity which is vital for FogAPs, unlike the weighted sum-rate optimizations in (\ref{eq:objCRAN}) or (\ref{eq:objFRAN}) which require the high processing capabilities of cloud BBUs. 

Thus, each FogAP $r$ solves the following optimization problem for each associated user $k \in \mathcal{K}_r$. 
Moreover, here we assume equal power allocation of the FogAP power among its associated users. Note that in the follow-up work, we will propose to reuse the pre-scheduling solution obtained from (\ref{eq:objFRAN}) for optimized power allocation as well.  
The optimization problem is thus

\begin{equation}
\max_{\w_{rk}} \zeta_{rk} \quad \text{s.t.}  \quad ||{\w_{rk}}||_2^2 \le \frac{P_r}{K_r},
\label{eq:objZeta}
\end{equation}
for which the closed-form solution is given in \cite{Sad07apr} 
\small
\begin{equation}
	\w_{rk}^{opt} = \sqrt{\frac{P_r}{K_r}}\max\mathrm{eig} \left\{\left(\sum_{k'\neq k}\h_{rk'}\h_{rk'}^H + \frac{K_r \sigma_n^2}{P_r}\I\right)^{-1}\h_{rk}\h_{rk}^H\right\},
\label{eq:beamformerSolution}
\end{equation}
\normalsize
where $\max\mathrm{eig}(A)$ gives the eigenvector corresponding to the largest eigenvalue of matrix $A$. 


%

\section{Numerical Results}
\label{sec:results}


We consider a 7-cell wrap-around two-tier CRAN/FogRAN to evaluate the reference and proposed algorithms. There are 3 macro-RRHs/FogAPs and 9 pico-RRHs/FogAPs, where the transmit power of macro and pico-RRHs are 43 and 30 dBm respectively, and their fronthaul capacities $C_r$ are set to (690,107) Mbps as in \cite{Dai15sep}. All channels are subject to Rayleigh fading and log-normal shadowing. The noise power spectral density is assumed to be $\sigma_n^2=-169$dBm/Hz, and the bandwidth 10 MHz. Other system parameters also follow that of \cite{Dai15sep}.

\begin{figure}[t]
\centering
\includegraphics[scale=0.3]{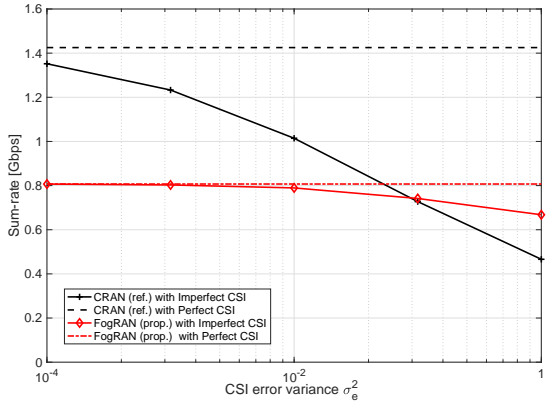} 
\vspace*{-0.5cm}
\caption{Sum-rate performance of reference CRAN and proposed Fog RAN algorithms, different levels of CSI imperfectness}
\label{fig:SumRate}
\end{figure}

In Fig. \ref{fig:SumRate}, we evaluate the system sum-rate for $K=60$ users for the reference centralized weighted sum-rate optimization for CRAN in Section \ref{sec:reference}, denoted \emph{CRAN (ref.)}, and the proposed pre-scheduling and local beamforming algorithm for FogRAN in Section \ref{sec:proposed}, denoted \emph{FogRAN (prop.)}. For the proposed method, the pre-scheduling period was fixed to $T=10$. Fig. \ref{fig:SumRate} shows the sum-rate degradation for different levels of CSI imperfectness given by the CSI error variance $\sigma_e^2$ defined in Section \ref{subsec:imperfectCSI} compared to the perfect CSI case. Clearly, the centralized algorithm offers very high throughput for near-perfect CSI, but degrades rapidly as the error variance grows. By contrast, the proposed algorithm for FogRAN shows a throughput loss due to the distributed beamforming for high quality CSI, but also a much higher robustness against CSI errors and even a close to optimal performance for $\sigma_e^2=1$. For realistic levels of CSI imperfectness where $\sigma_e^2 \geq 0.1$ as pointed out in  \cite{Du12mar}, our proposed algorithm for FogRAN even outperforms the centralized reference algorithm. 

\begin{figure}[t]
\centering
\includegraphics[scale=0.3]{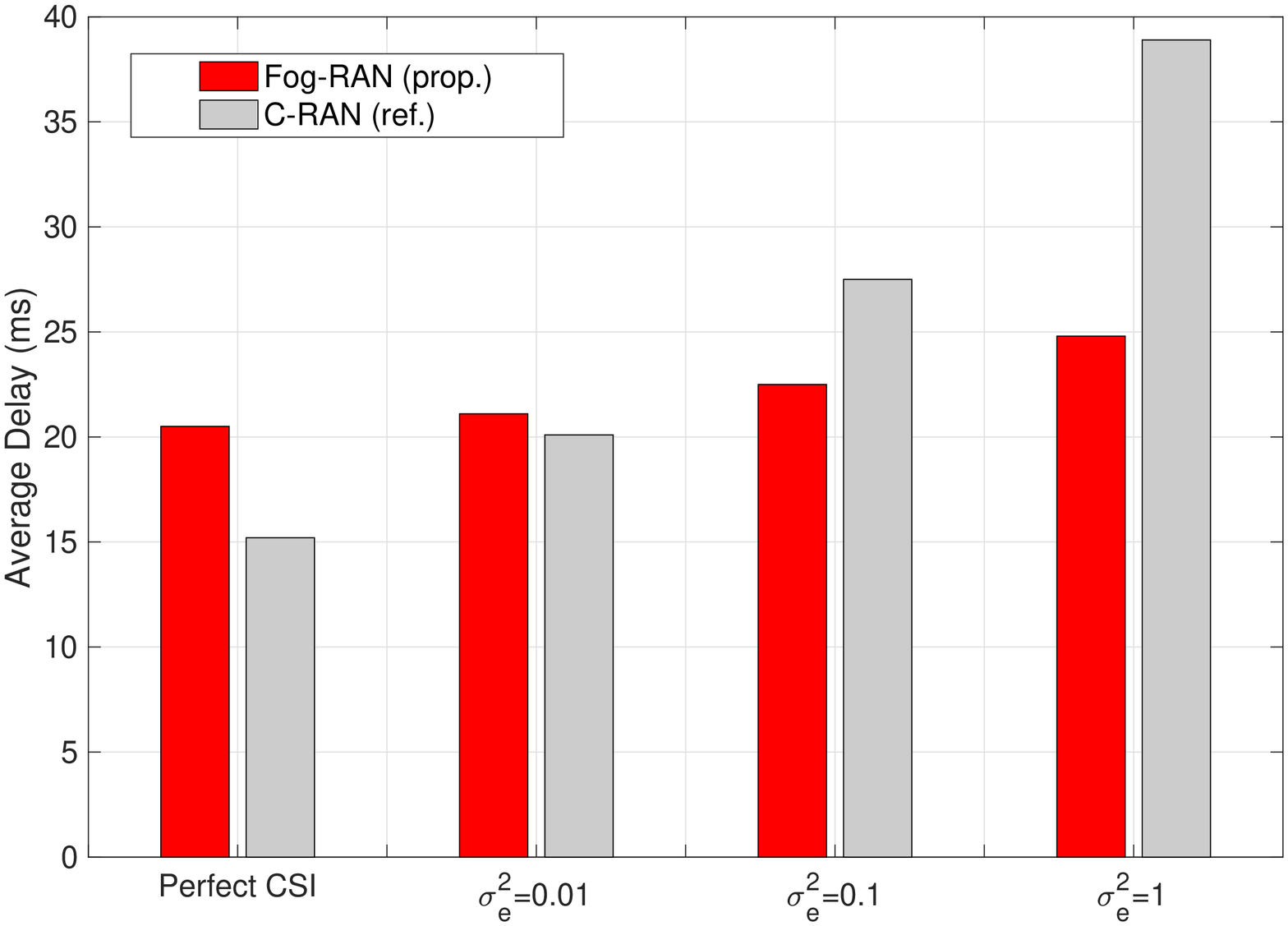} 
\vspace*{-0.6cm}
\caption{Delay performance of reference and proposed algorithms, different levels of CSI imperfectness, 12 kbits packet}
\label{fig:Delay1}
\end{figure}

\begin{figure}[t]
\centering
\includegraphics[scale=0.3]{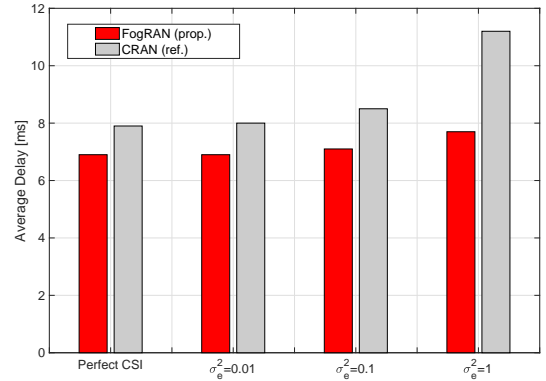} 
\vspace*{-0.6cm}
\caption{Delay performance of reference and proposed algorithms, different levels of CSI imperfectness, 1 kbit packet}
\label{fig:Delay2}
\end{figure}

Next, we evaluate the average delay performance of these algorithms, one of the main motivations for FogRANs and edge computation. The delay is defined as the time required for receiving a packet of $P$ bits, averaged over all users. 
Fig. \ref{fig:Delay1} shows the delay performance for a relatively large packet size of $12$ kbits. The reference centralized algorithm for CRAN is better for perfect and near-perfect CSI, but is outperformed by the proposed scheme as the CSI error grows. However, for a smaller packet size of $1$ kbit in Fig. \ref{fig:Delay2}, the proposed algorithm always outperforms the reference one, even for perfect CSI. This is because even though the proposed scheme achieves lower total throughput, it enables to serve high enough rates through the distributed but accurate beamforming, so that small packets are received efficiently. On the contrary, the centralized scheme allows to boost the throughput by globally concentrating the resources towards the users with best channel conditions, at the detriment of users in lower conditions. But the throughput of the best users diminishes drastically as the CSI errors increase, thereby degrading the delay performance. Observing all figures, we can conclude that the proposed scheme allows to improve the system throughput and delays for large and small packets simultaneously, in the range of realistic CSI imperfectness.


\vspace*{-0.3cm}


\section{Conclusion}
\label{sec:conclude}

We proposed a hybrid semi-distributed resource allocation algorithm suited for FogRANs with centralized user pre-scheduling carried out periodically at cloud BBUs and distributed local beamforming at each FogAP in each frame. Although optimal, the centralized algorithm that jointly solves the user clustering and beamforming in CRANs can only make use of imperfect CSIs due to the inevitable transport delays on fronthaul links. 
Therefore, our algorithm takes advantage of both the large-scale cloud processing to optimize the user pre-scheduling despite imperfect CSIs, and the availability of perfect CSIs at FogAPs for accurate beamforming, despite local optimization.   
The simulation results show the effectiveness of the proposed method for realistic levels of imperfect CSI, both in terms of system throughput and delays. In particular, the delay improvements for small packets suggest that our approach is well-suited to support future IoT applications that typically generate a large amount of very small packets. 
  
This work has opened up key issues to investigate, among which the optimized design of pre-scheduling/beamforming and CSI acquisition under high user mobility.






\bibliographystyle{IEEEtran}
\bibliography{Refs_FogRAN}

\end{document}